\documentclass{llncs}

\usepackage{listings}
\usepackage{amssymb}
\usepackage{url}
\usepackage{ifthen}
\usepackage{graphicx}
\usepackage{verbatim}
\usepackage{hhline}
\usepackage{threeparttable}
\usepackage{placeins}

\newboolean{LONGVERSION}
\setboolean{LONGVERSION}{false}

\ifthenelse{\boolean{LONGVERSION}}{%%
\usepackage{varioref}
}{%%\ifthenelse{\boolean{LONGVERSION}}{%%
}%%\ifthenelse{\boolean{LONGVERSION}}{%%

\newcommand*{\Cplusplus}{{C\nolinebreak[4]\hspace{-.05em}\raisebox{.4ex}{\tiny\bf ++}}}

\begin{document}

\title{The MISRA C Coding Standard and its Role in the Development
       and Analysis of Safety- and Security-Critical Embedded Software}
\author{Roberto Bagnara\inst{1,2}\thanks{While Roberto Bagnara
    is a member of the \emph{MISRA C Working Group} and of
    ISO/IEC JTC1/SC22/WG14, a.k.a.\ the \emph{C Standardization Working Group},
    the views expressed in this paper are his and his coauthors' and should
    not be taken to represent the views of either working group.}
  \and Abramo Bagnara\inst{1}
  \and Patricia M. Hill\inst{1}
}
\authorrunning{R.~Bagnara,
               A.~Bagnara,
               P.~M.~Hill
}
\tocauthor{Roberto Bagnara (BUGSENG srl, University of Parma),
           Abramo Bagnara (BUGSENG srl)
           Patricia Hill (BUGSENG srl)
}

\institute{
    BUGSENG srl,
    \url{http://bugseng.com} \\
    \email{name.surname@bugseng.com}
    \and
    Department of Mathematical, Physical and Computer Sciences,
    University of Parma,
    Italy \\
    \email{bagnara@cs.unipr.it}
}

\maketitle

\begin{abstract}
The MISRA project started in 1990 with the mission of providing
world-leading best practice guidelines for the safe and secure
application of both embedded control systems and standalone software.
MISRA C is a coding standard defining a subset of the C language,
initially targeted at the automotive sector, but now adopted across
all industry sectors that develop C software in safety- and/or
security-critical contexts.  In this paper, we introduce MISRA C,
its role in the development of critical software, especially in
embedded systems, its relevance to industry safety standards, as well as
the challenges of working with a general-purpose programming language
standard that is written in natural language with a slow evolution
over the last 40+ years.  We also outline the role of static analysis
in the automatic checking of compliance with respect to MISRA C, and
the role of the MISRA C language subset in enabling a wider application
of formal methods to industrial software written in C.
\end{abstract}

\section{Introduction}
\label{sec:introduction}

In September 1994, the ``First International Static Analysis Symposium''
took place in Namur, Belgium \cite{SAS94}.  The \emph{Call for Papers}
contained the following:
\begin{quote}
Static Analysis is increasingly recognized as a fundamental tool for
high performance implementations and verification systems of high-level
programming languages. The last two decades have witnessed substantial
developments in this area, ranging from the theoretical frameworks to
the design and implementation of analysers and their applications in
optimizing compilers.
\end{quote}

In November 1994, MISRA\footnote{Originally, an acronym
  for \emph{Motor Industry Software Reliability Association}.}
published its ``Development Guidelines For Vehicle Based Software''
\cite{MISRA-1994}.  These listed static analysis as the first automatic
methodology to verify software and contained the following paragraphs:
\begin{quote}
\begin{description}
\item[3.5.1.5]
Consideration should be given to using a restricted subset of a
programming language to aid clarity, assist verification and
facilitate static analysis where appropriate.
\item[3.5.2.6]
Static analysis is effective in demonstrating that a program is well
structured with respect to its control, data and information flow. It
can also assist in assessing its functional consistency with its
specification.
\end{description}
\end{quote}
Paragraph 3.5.1.5 led to the definition of the subset of the C~programming
language that will later be called \emph{MISRA~C}.

While the quoted texts show the passage of time (today we would express
things differently), they witness the fact that static analysis
was recognized as an established research field at about the
same time that it gathered enough industrial recognition to be
explicitly recommended by an influential set of guidelines for
the automotive industry, one of the most important economic
sectors by revenue.
The connection between static analysis research
and the industrial world ---which now encompasses all industry sectors---
that recognizes MISRA~C as the basis for the development of safe
and secure applications in C has been basically unidirectional
and mediated by the tool providers.
These tool providers are interested in
all advances in static analysis research in order to improve the
applicability and usefulness of their tools and hence simplify
the task of verifying compliance with respect to the MISRA~C guidelines.
It must be admitted that the static analysis research community has
seen the (very pragmatic) industry movement behind MISRA~C with
a somewhat snobbish and often not well informed
attitude.\footnote{The authors of this paper are \emph{not}
an exception to this statement, at least not until 2010.}
For instance, \cite[Section 3]{CousotCFMMMR07} suggests that MISRA~C
concerns coding style and that semantic-based static analysis is not needed
to check its guidelines.
In reality, while MISRA~C encourages the adoption of a consistent
programming style, it has always left this matter to individual
organizations: ``In addition to adopting the subset, an organisation should
also have an in-house style guide. [\dots]
However the enforcement of the style guide is outside the scope
of this document'' \cite[Section~4.2.2]{MISRA-C-2004} (see also
\cite[Section 5.2.2]{MISRA-C-2012}).
Moreover, as we will see, semantic-based static analysis is required
to check many MISRA~C guidelines without constraining too much
the way the code is written.

In this paper we try to clear up such misconceptions
and to properly introduce MISRA~C to the static analysis community.
Our ultimate aim is to foster collaboration between the communities,
one which we believe could be very fruitful: the wide
adoption of MISRA~C in industry constitutes an avenue for a wider
introduction of formal methods and a good opportunity to channel
some applied static analysis research to the most important subset
of the C programming language.

The plan of the paper is the following:
Section~\ref{sec:c-language} introduces the C~language
explaining why it is so widely adopted, why it is (not completely)
defined as it is, why it is not going to change substantially any
time soon, and why subsetting it is required;
Section~\ref{sec:misra-c} introduces the MISRA project and MISRA~C
focusing on its last edition, MISRA~C:2012, with its amendments
and addenda;
Section~\ref{sec:static-analysis-and-misra-c} highlights the links
between MISRA~C and static analysis;
Section~\ref{sec:discussion} discusses some trends and opportunities;
Section~\ref{sec:conclusion} concludes.

\section{The C Language}
\label{sec:c-language}

The development of the C~programming language started in 1969 at Bell
Labs, almost 50 years ago, and the language was used for the development
of the Unix operating system \cite{Ritchie93}.
Despite frequent criticism, C is still one of the most used programming
languages overall\footnote{%
  Source: TIOBE Index for June 2018,
  see \url{https://www.tiobe.com/tiobe-index/}.}
and the most used one for the development of embedded systems
\cite{BarrGroupSurvey2018,VDCSurvey2011}.
There are several reasons why C has been and is so successful:
\begin{itemize}
\item
  C compilers exist for almost any processor, from tiny DSPs used in
  hearing aids to supercomputers.
\item
  C compiled code can be very efficient and without hidden costs,
  i.e., programmers can roughly predict running times even before
  testing and before using tools for worst-case execution time
  approximation.\footnote{This is still true for implementations
    running on simple processors, with a limited degree of caching
    and internal parallelism.  Prediction of maximum running time
    without tools becomes outright impossible for current multi-core
    designs such as Kalray MPPA, Freescale P4080, or ARM Cortex-A57
    equivalents (see, e.g., \cite{NelisYP16,NowotschP12,NowotschPBTWS14}).}
\item
  C allows writing compact code: it is characterized by the availability
  of many built-in operators, limited verbosity, \dots
\item
  C is defined by international standards:
  it was first standardized in 1989 by the
  American National Standards Institute (this version of the language
  is known as ANSI~C) and then by the International Organization
  for Standardization (ISO)
  \cite{ISO-C-1990,ISO-C-1995,ISO-C-1999,ISO-C-1999-consolidated-TC3,ISO-C-2011}.
\item
  C, possibly with extensions, allows easy access to the hardware and
  this is a must for the development of embedded software.
\item
  C has a long history of usage in all kinds of systems including
  safety-, security-, mission- and business-critical systems.
\item
  C is widely supported by all sorts of tools.
\end{itemize}
Claims that C would eventually be superseded by \Cplusplus{} do not
seem very plausible, at least as far as the embedded software industry
is concerned.  In addition to the already-stated motives, there is language
size and stability: \Cplusplus{} has become a huge, very complex language;
moreover it is evolving at a pace that is in sharp contrast with industrial
best practices.  The trend whereby \Cplusplus{} rapid evolution
clashes with the industry requirements for stability and backward
compatibility has been put black-on-white at a recent
WG21 meeting,\footnote{%
  WG21 is a common shorthand for ISO/IEC JTC1/SC22/WG21, a.k.a.\ the
  \Cplusplus{} \emph{Standardization Working Group}.
  The cited meeting took place in Jacksonville, FL, USA,
  March 12--17, 2018.}
where the following statement was agreed upon \cite{Winters18}:
``The Committee should be willing to consider the design / quality of
proposals even if they may cause a change in behavior or failure to
compile for existing code.''

A good portion of the criticism of C comes from the notion of
\emph{behavior}, defined as \emph{external appearance or action}
\cite[Par.~3.4]{ISO-C-1999-consolidated-TC3} and the so-called \emph{as-if rule},
whereby the compiler is allowed to do any transformation that ensures that the
``observable behavior'' of the program is the one described by the
standard
\cite[Par~5.1.2.3\#5]{ISO-C-1999-consolidated-TC3}.\footnote{In this paper,
  we refer to the C99 language standard \cite{ISO-C-1999} because
  this is the most
  recent version of the language that is targeted by the current
  version of MISRA~C \cite{MISRA-C-2012}.
  All what is said about the C language itself
  applies equally, with only minor variations,
  to all the published versions of the C standard.}
While all compiled languages have a sort of \emph{as-if rule} that
allows optimized compilation, one peculiarity of C is that it is
not fully defined.  There are four classes of not fully defined behaviors
(in the sequel, collectively referred to as ``non-definite behaviors''):
\begin{description}
\item[implementation-defined behavior:]
\emph{unspecified behavior where each implementation documents how the choice is made}
\cite[Par.~3.4.1]{ISO-C-1999-consolidated-TC3}; e.g., the sizes and precise representations
of the standard integer types;
\item[locale-specific behavior:]
\emph{behavior that depends on local conventions of nationality, culture, and
language that each implementation documents}
\cite[Par.~3.4.2]{ISO-C-1999-consolidated-TC3}; e.g., character sets and how characters
are displayed;
\item[undefined behavior:]
\emph{behavior, upon use of a non-portable or erroneous program construct or
of erroneous data, for which this International Standard imposes
no requirements}
\cite[Par.~3.4.3]{ISO-C-1999-consolidated-TC3}; e.g., attempting to
write a string literal constant or shifting an expression
by a negative number or by an amount greater than or equal to the width
of the promoted expression;
\item[unspecified behavior:]
\emph{use of an unspecified value, or other behavior where this
International Standard provides two or more possibilities and imposes
no further requirements on which is chosen in any instance}
\cite[Par.~3.4.4]{ISO-C-1999-consolidated-TC3}; e.g., the order in which
sub-expressions are evaluated.
\end{description}

Setting aside locale-specific behavior, whose aim is to avoid some
nontechnical obstacles to adoption,
it is important to understand the connection between non-definite
behavior and the relative ease with which optimizing compilers
can be written.  In particular, C data types and operations can be
directly mapped to data types and operations of the target
machine.  This is the reason why the sizes and precise representations
of the standard integer types are implementation-defined: the implementation
will define them in the most efficient way depending on properties
of the target CPU registers, ALUs and memory hierarchy.
Attempting to write on string literal constants is undefined behavior
because they may reside in read-only memory and/or may be merged and shared:
for example, a program containing \verb|"String"| and \verb|"OtherString"|
may only store the latter and use a suffix of that representation to represent
the former.
The reason why shifting an expression by a negative number
or by an amount greater than or equal to the width of the promoted expression
is undefined behavior is less obvious.
What sensible semantics can be assigned
to shifting by a negative number of bit positions?  Shifting in the opposite
direction is a possible answer, but this is usually not supported in hardware,
so it would require a test, a jump and a negation.
It is a bit more subtle to understand why the following is undefined behavior:
\begin{lstlisting}
  uint32_t i = 1;
  i = i << 32;  /* Undefined behavior. */
\end{lstlisting}
One would think that pushing $32$ or more zeroes to the right
of \verb|i| would give zero.
However, this does not correspond to how some architectures implement
shift instructions.  IA-32, for instance
\cite[section on ``IA-32 Architecture Compatibility'']{Intel18}:
\begin{quote}
The 8086 does not mask the shift count. However, all other IA-32 processors
(starting with the Intel 286 processor) do mask the shift count to 5 bits,
resulting in a maximum count of 31. This masking is done in all operating modes
(including the virtual-8086 mode) to reduce the maximum execution time of the
instructions.
\end{quote}
This means that, on all IA-32 processors starting with the Intel 286,
a direct mapping of C's right shift to the corresponding machine instruction
will give:
\begin{lstlisting}
  i = i << 32;           /* This is equivalent to... */
  i = i << (32 & 0x1F);  /* ... this, i.e., ...      */
  i = i << 0;            /* this, which is a no-op.  */
\end{lstlisting}
So also for this case, for speed and ease of implementation,
C leaves the behavior undefined.

The recurring request to WG14\footnote{%
  Short for ISO/IEC JTC1/SC22/WG14,
  a.k.a. the \emph{C Standardization Working Group}.}
to ``fix the language'' is off the mark.
In fact, weakness of the C language comes from its strength:
\begin{itemize}
 \item
Non-definite behavior is the consequence of two factors:
\begin{enumerate}
\item
the ease of writing efficient compilers for almost any architecture;
\item
the existence of many compilers by different vendors
and the fact that the language is standardized.
\end{enumerate}
Concerning the second point, it should be considered that,
in general, ISO standardizes existing practice
taking into account the opinions of the vendors that participate
in the standardization process, and with great attention
to backward compatibility:
so, when diverging implementations exist, non-definite behavior might
be the only way forward.
\item
The objective of easily obtaining efficient code with no hidden costs
is the reason why, in C, there is no run-time error checking.
\item
Easy access to the hardware entails the facility with which
the program state can be corrupted.
\item
Code compactness is one of the reasons why the language can easily be
misunderstood and misused.
\end{itemize}

Summarizing, the C language can be expected to remain faithful
to its original spirit and to be around for the foreseeable future,
at least for the development of embedded systems.
However, it is true that several features of C do conflict with
both safety and security requirements.
For this reason, \emph{language subsetting} is crucial for critical
applications.
This was recognized early in \cite{Hatton95} and is now
mandated or recommended by all safety- and
security-related industrial standards, such as
IEC 61508 (industrial),
ISO~26262 (automotive), CENELEC~EN~50128 (railways),
RTCA DO-178B/C (aerospace) and
FDA's \emph{General Principles of Software Validation} \cite{FDA02}.
Today, the most authoritative language subset for the C programming language
is \emph{MISRA~C}, which is the subject of the next section.

\section{MISRA C}
\label{sec:misra-c}

The MISRA
project started in 1990 with the mission of providing
world-leading best practice guidelines for the safe and secure
application of both embedded control systems and standalone software.
The original project was part of the UK Government’s ``SafeIT'' programme
but it later became self-supporting, with MIRA Ltd, now HORIBA MIRA Ltd,
providing the project management support.
Among the activities of MISRA is the development of guidance in specific
technical areas, such as the C and \Cplusplus{} programming languages,
model-based development and automatic code generation,
software readiness for production, safety analysis, safety cases and so on.
In November 1994, MISRA published its
``Development guidelines for vehicle based software'',
a.k.a.\ ``The MISRA Guidelines'' \cite{MISRA-1994}:
this is the \emph{first} automotive publication concerning functional safety,
more than 10 years before work started on ISO 26262 \cite{ISO-26262:2011}.

The MISRA guidelines \cite{MISRA-1994} prescribed the use of
``a restricted subset of a standardized structured language.''
In response to that, the MISRA consortium began work on
the MISRA C guidelines: at that time Ford and Land Rover
were independently developing in-house rules for vehicle-based
C software and it was recognized that a common activity would
be more beneficial to industry.
The first version of the MISRA~C guidelines was published
in 1998~\cite{MISRA-C-1998} and received significant industrial
attention.

In 2004, following the many comments received from its users
---many of which, beyond expectation, were in non-automotive industries---
MISRA published an improved version of the C guidelines \cite{MISRA-C-2004}.
In MISRA~C:2004 the intended audience explicitly became constituted by
\emph{all} industries that develop C software for use in high-integrity/critical
systems.
Due to the success of MISRA~C and the fact that \Cplusplus{}
is also used in critical contexts, in 2008 MISRA published  a similar set
of \emph{MISRA~\Cplusplus} guidelines~\cite{MISRA-CPP-2008}.

\begin{figure}
\begin{center}
\includegraphics[width=12cm]{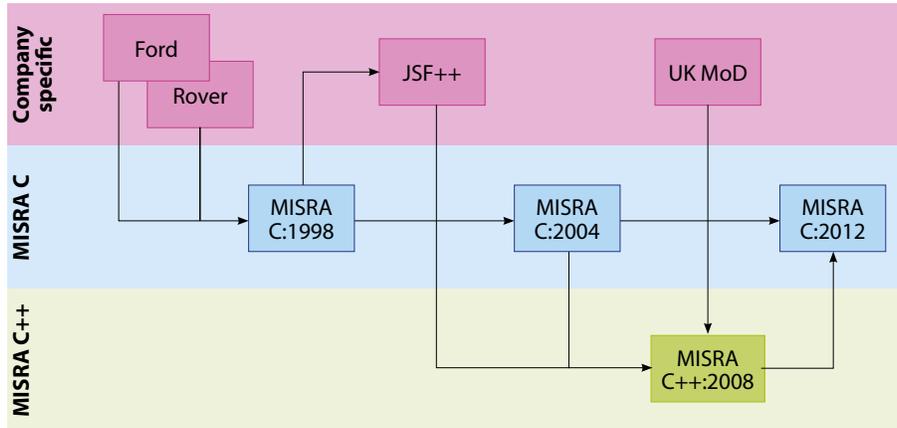}
\end{center}
\caption{Origin and history of MISRA C}
\label{fig:misra-c-history}
\end{figure}

Both MISRA~C:1998 and MISRA~C:2004 target the 1990 version of the
C Standard~\cite{ISO-C-1990}.
In 2013, the revised set of guidelines known as MISRA~C:2012
was published~\cite{MISRA-C-2012}.
In this version
there is support both for C99 \cite{ISO-C-1999} as well as C90
(in its amended and corrected form sometimes referred to as C95
\cite{ISO-C-1995}).
With respect to previous versions, MISRA~C:2012 covers more language
issues and provides a more precise specification of the guidelines
with improved rationales and examples.
Figure~\ref{fig:misra-c-history} shows part of the relationship
and influence between the MISRA C/\Cplusplus{} guidelines and other sets of
guidelines.  It can be seen that MISRA~C:1998 influenced
Lockheed's ``JSF Air Vehicle \Cplusplus{} Coding Standards for
the System Development and Demonstration Program'' \cite{JSF-CPP-2005},
which influenced
MISRA~\Cplusplus{}:2008, which, in turn, influenced MISRA~C:2012.
The activity that led to MISRA~\Cplusplus{}:2008 was also encouraged
by the UK Ministry of Defence which, as part of its
\emph{Scientific Research Program}, funded a
work package that resulted in the development of a ``vulnerabilities document''
(the equivalent of Annex~J listing the various behaviors in ISO~C,
which is missing in ISO~\Cplusplus{}, making it hard work to identify them
and to ensure they are covered by the guidelines).
Moreover, MISRA~C deeply influenced NASA's
``JPL Institutional Coding Standard for the C Programming Language''
\cite{NASA-JPL-C-2009} and several other coding standards (see, e.g.,
\cite{BARR-C-2013,CERT-C-2016,IPA-ESCR-C-2014}).

The MISRA C guidelines are concerned with aspects of C that impact on
the safety and security of the systems, whether embedded or standalone:
they define ``a subset of the C language in which the opportunity to
make mistakes is either removed or reduced'' \cite{MISRA-C-2012}.
The guidelines ban critical non-definite behavior and constrain
the use of implementation-defined behavior and compiler extensions.
They also limit the use of language features that can easily be
misused or misunderstood.  Overall, the guidelines are designed to
improve reliability, readability, portability and maintainability.

There are two kinds of MISRA C guidelines.
\begin{description}
\item[Directive:]
a guideline where the information concerning compliance
is generally not fully contained in the source code:
requirements, specifications, design, \dots\,
may have to be taken into account.
Static analysis tools may assist in checking compliance with respect
to directives if provided with extra information not derivable from
the source code.
\item[Rule:]
a guideline such that information concerning compliance
is fully contained in the source code.
Discounting undecidability, static analysis tools should, in principle,
be capable of checking compliance with respect to the rule.
\end{description}

A crucial aspect of MISRA~C is that it has been designed to be used
within the framework of a documented development process where
justifiable non-compliances will be authorized and recorded
as \emph{deviations}.
To facilitate this, each MISRA C guideline has been assigned a category.
\begin{description}
\item[Mandatory:]
C code that complies to MISRA C must comply with every
man\-datory guideline; deviation is not permitted.
\item[Required:]
C code that complies to MISRA C shall comply with every
required guideline; a formal deviation is required
where this is not the case.
\item[Advisory:]
these are recommendations that should be followed as far as is
reasonably practical;
formal deviation is not required, but non-compliances should be documented.
\end{description}
Every organization or project may choose to treat any required guideline
as if it were mandatory and any advisory guideline as if it were required
or mandatory.
The adoption of MISRA~Compliance:2016 \cite{MISRA-Compliance-2016}
allows advisory guidelines to be downgraded to ``Disapplied'' when a check
for compliance is considered to have no value, e.g., in the case of
\emph{adopted code}\footnote{Such as the standard library,
device drivers supplied by the compiler vendor or the hardware manufacturer,
middleware components,
third party libraries,
automatically generated code,
legacy code, \dots}
that has not been developed so as to comply with the MISRA~C guidelines.
Of course, the decision to disapply a guideline should not be taken lightly:
\cite{MISRA-Compliance-2016} prescribes the compilation of a
\emph{guideline recategorization plan} that must contain, among other
things, the rationale for any decision to disapply a guideline.

Each MISRA~C rule is marked as \emph{decidable} or \emph{undecidable}
according to whether answering the question ``Does this code comply?''
can be done algorithmically.
Hence rules are marked `undecidable' whenever violations depend on
run-time (dynamic) properties such as the value contained in a modifiable
object or whether control reaches a particular point.
Conversely, rules are marked `decidable'
whenever violations depend only on compile-time (static) properties,
such as the types of the objects or the names and the scopes of identifiers.
Clearly, for rules marked `decidable',
it is theoretically possible (i.e., given
adequate computational resources) for a tool to emit a message
if and only if the rule is violated.
However, for rules marked `undecidable', any tool will have to deal with
the \emph{don't know} answer in addition to \emph{yes} and \emph{no}
at each distinct, relevant program point.
In either case, if it is not practical (or even possible)
for the tool to decide if the code is compliant with respect
to a guideline at a particular program point, it can:
\begin{itemize}
\item
suppress the \emph{don't know} answer
  (i.e., possibly false negatives, no false positives);
\item
emit the \emph{don't know} answer as a \emph{yes} message
  (i.e., no false negatives, possibly false positives);
\item
a combination of the above
  (i.e., both possibly false negatives and possibly false positives);
\item
emit the \emph{don't know} answer as a \emph{caution} message
  (i.e., no false negatives, confined, possibly false positives).
\end{itemize}

MISRA C rules are also classified according to the amount of code
that needs to be analyzed in order to detect all violations of the rule.
\begin{description}
\item[Single Translation Unit:]
all violations within a project can be detected by checking
each translation unit independently.
\item[System:]
identifying violations of a rule within a translation unit requires
checking more than the translation unit in question, if not all
the source code that constitutes the system.
\end{description}

MISRA~C:2012 Amendment 1 \cite{MISRA-C-2012-Amendment1}, published in
2016, enhances MISRA C:2012 so as to extend its applicability to
industries and applications where data-security is an issue.
It includes $14$ new guidelines ($1$ directive and $13$ rules)
to complete the coverage of
ISO/IEC TS 17961:2013 \cite{ISO-IEC_DTS_17961-2013},
a.k.a. \emph{C Secure Coding Rules}, a set of rules
for secure coding in~C.\footnote{This technical specification has been
slightly amended in 2016 \cite{ISO-IEC_DTS_17961-2016}.}
Details of such complete coverage are provided in
\cite{MISRA-C-2012-Addendum-2-2}.
A similar document \cite{MISRA-C-2012-Addendum-3} shows that,
with Amendment 1, coverage of \emph{CERT C Coding Standard} is almost
complete and that, consequently, MISRA~C is today the language
subset of choice for all industries developing embedded systems
in C that are safety- and/or security-critical
\cite{Bagnara16SPIN}.

For the rest of this paper, all references to \emph{MISRA~C}
will be for the latest published version MISRA~C:2012 \cite{MISRA-C-2012}
including its Technical Corrigendum~1 \cite{MISRA-C-2012-TC1} and
Amendment~1 \cite{MISRA-C-2012-Amendment1}: these will be consolidated
into the forthcoming first revision of MISRA~C:2012 \cite{Banks16}.
It should be noted that both the MISRA~C and MISRA~\Cplusplus{} projects
are active and constantly improving the guidelines and developing new works:
for instance, the MISRA~C Working Group is currently working at adding support
for C11 \cite{ISO-C-2011} and, in response to community feedback,
at further enhancing the guidance on undefined/unspecified behaviors
\cite{Banks16}.

\section{Static Analysis and MISRA C}
\label{sec:static-analysis-and-misra-c}

The majority of the MISRA~C guidelines are decidable, and thus
compliance can be checked by algorithms that:
\begin{itemize}
\item
  do not need nontrivial approximations of the value of program objects;
\item
  do not need nontrivial control-flow information.
\end{itemize}
Of course, these algorithms can still be very complex.
For instance, the nature of the translation process of the C~language,
which includes a preprocessing phase, is a source of complications:
the preprocessing phase must be tracked precisely, and compliance
may depend on the source code before preprocessing, on the source code after
preprocessing, or on the relationship between the source code before and
after preprocessing.

The rest of this section focuses on those guidelines whose check for compliance
requires or significantly benefits from semantic-based analysis.
Obviously every undecidable rule has decidable approximations,
but these are necessarily characterized by a significant number
of false positives unless rigid programming schemes are adopted.
For example, Rule~17.2, which disallows recursion, admits a decidable
approximation that requires finding cycles in the call graph and
flagging, as potentially non-compliant, all function calls via pointers.
If function calls via pointers are not used (i.e., the program is written
in a smaller subset of C than that strictly mandated by the rule) then there
will be no false positives.

{
\setlength{\tabcolsep}{2pt}
\renewcommand{\arraystretch}{1.13}
\begin{table}[p]
\centering
\caption{MISRA~C guidelines whose checking requires/benefits from semantic analysis}
\label{tab:MISRA-C-semantic-guidelines}
\begin{tabular}{l||l}
Guideline & Rough one-line summary \\
\hhline{=::=}
\textbf{D4.1} &
{Avoid run-time failures} \\
\hhline{-||-}
\textbf{D4.11} &
{Check the validity of values passed to library functions} \\
\hhline{-||-}
\textbf{D4.13} &
{Resource-handling functions should be called in an appropriate sequence} \\
\hhline{-||-}
\textbf{D4.14} &
{Do not trust values received from external sources} \\
\hhline{-||-}
\textbf{R1.3} &
{No undefined or critical unspecified behavior} \\
\hhline{-||-}
\textbf{R2.1} &
{No unreachable code} \\
\hhline{-||-}
\textbf{R2.2} &
{No dead code} \\
\hhline{-||-}
\textbf{R8.13} &
{Point to \texttt{const}-qualified type if possible} \\
\hhline{-||-}
\textbf{R9.1} &
{Do not read uninitialized automatic storage} \\
\hhline{-||-}
\textbf{R12.2} &
{Right-hand operand of a shift operator must be in range} \\
\hhline{-||-}
\textbf{R13.1} &
{No side effects in initializers} \\
\hhline{-||-}
\textbf{R13.2} &
{Do not depend on unspecified evaluation order of expressions} \\
\hhline{-||-}
\textbf{R13.5} &
{No side effects in right-hand operand of \texttt{\&\&} or \texttt{||}} \\
\hhline{-||-}
\textbf{R14.1} &
{No floating-point loop counters} \\
\hhline{-||-}
\textbf{R14.2} &
{Restricted form of \texttt{for} loops} \\
\hhline{-||-}
\textbf{R14.3} &
{No invariant controlling expressions} \\
\hhline{-||-}
\textbf{R17.2} &
{No direct or indirect recursion} \\
\hhline{-||-}
\textbf{R17.5} &
{Actual parameters for arrays must have an appropriate size} \\
\hhline{-||-}
\textbf{R17.8} &
{Do not modify function parameters} \\
\hhline{-||-}
\textbf{R18.1} &
{Pointer arithmetic must not exceed array limits} \\
\hhline{-||-}
\textbf{R18.2} &
{Do not subtract pointers not pointing to the same array} \\
\hhline{-||-}
\textbf{R18.3} &
{Do not compare pointers not pointing to the same object} \\
\hhline{-||-}
\textbf{R18.6} &
{Pointer object must not live longer than corresponding pointees} \\
\hhline{-||-}
\textbf{R19.1} &
{Objects must not be assigned or copied to overlapping objects} \\
\hhline{-||-}
\textbf{R21.13} &
{Functions in \texttt{<ctype.h>} must not be passed out-of-spec values} \\
\hhline{-||-}
\textbf{R21.14} &
{Do not use \texttt{memcmp} to compare null-terminated strings} \\
\hhline{-||-}
\textbf{R21.17} &
{Use of functions from \texttt{<string.h>} must not result in buffer overflow} \\
\hhline{-||-}
\textbf{R21.18} &
{\texttt{size\_t} argument of functions from \texttt{<string.h>} must be in range} \\
\hhline{-||-}
\textbf{R21.19} &
{Do not modify objects through pointers returned by \texttt{localeconv}, \dots} \\
\hhline{-||-}
\textbf{R21.20} &
{Pointers returned by \texttt{asctime}, \texttt{ctime}, \dots\ must not be reused} \\
\hhline{-||-}
\textbf{R22.1} &
{All dynamically-obtained resources must be explicitly released} \\
\hhline{-||-}
\textbf{R22.2} &
{Do not free memory that was not dynamically allocated} \\
\hhline{-||-}
\textbf{R22.3} &
{Do not open files for read and write at the same time on different streams} \\
\hhline{-||-}
\textbf{R22.4} &
{Do not attempt to write to a read-only stream} \\
\hhline{-||-}
\textbf{R22.5} &
{Do not directly access the content of a \texttt{FILE} object} \\
\hhline{-||-}
\textbf{R22.6} &
{Do not use the value of pointer to a \texttt{FILE} after the stream is closed} \\
\hhline{-||-}
\textbf{R22.7} &
{Macro \texttt{EOF} must only be compared to values returned by some functions} \\
\hhline{-||-}
\textbf{R22.8} &
{Reset \texttt{errno} before calling an \emph{errno-setting-function}} \\
\hhline{-||-}
\textbf{R22.9} &
{Test \texttt{errno} after calling an \emph{errno-setting-function}} \\
\hhline{-||-}
\textbf{R22.10} &
{Test \texttt{errno} only after calling an \emph{errno-setting-function}} \\
\hhline{-||-}
\end{tabular}
\end{table}
}

{
\setlength{\tabcolsep}{4pt}
\renewcommand{\arraystretch}{1}
\begin{table}[p]
\centering
\begin{threeparttable}
\caption{MISRA~C guidelines and main static analysis properties}
\label{tab:MISRA-C-semantic-guideline-properties}
\begin{tabular}{l||c||c|c}
Guideline & control-flow & \multicolumn{2}{c}{data-flow} \\
\hline
\multicolumn{2}{c||}{}       & points-to  & arithmetic \\
\hhline{=:t:=::=:=}
\textbf{D4.1}   &            & \checkmark & \checkmark \\
\hhline{-||-||-|-}
\textbf{D4.11}   &           & \checkmark & \checkmark \\
\hhline{-||-||-|-}
\textbf{D4.13}  & \checkmark & \checkmark &            \\
\hhline{-||-||-|-}
\textbf{D4.14}   &           &            & \checkmark \\
\hhline{-||-||-|-}
\textbf{R1.3}   & \checkmark & \checkmark & \checkmark \\
\hhline{-||-||-|-}
\textbf{R2.1}   & \checkmark &            &            \\
\hhline{-||-||-|-}
\textbf{R2.2}   & \checkmark &            & \checkmark \\
\hhline{-||-||-|-}
\textbf{R8.13}  &            &            & \checkmark \\
\hhline{-||-||-|-}
\textbf{R9.1}   & \checkmark\tnote{1}
                                          & \checkmark\tnote{1}
                                                       \\
\hhline{-||-||-|-}
\textbf{R12.2}  &            &            & \checkmark \\
\hhline{-||-||-|-}
\textbf{R13.1}  &            & \checkmark &            \\
\hhline{-||-||-|-}
\textbf{R13.2}  & \checkmark &            &            \\
\hhline{-||-||-|-}
\textbf{R13.5}  &            & \checkmark &            \\
\hhline{-||-||-|-}
\textbf{R14.1}  & \checkmark & \checkmark & \checkmark \\
\hhline{-||-||-|-}
\textbf{R14.2}  &            & \checkmark & \checkmark \\
\hhline{-||-||-|-}
\textbf{R14.3}  &            & \checkmark & \checkmark \\
\hhline{-||-||-|-}
\textbf{R17.2}  & \checkmark & \checkmark &            \\
\hhline{-||-||-|-}
\textbf{R17.5}  &            & \checkmark & \checkmark \\
\hhline{-||-||-|-}
\textbf{R17.8}  &            & \checkmark &            \\
\hhline{-||-||-|-}
\textbf{R18.1}  &            & \checkmark & \checkmark \\
\hhline{-||-||-|-}
\textbf{R18.2}  &            & \checkmark &            \\
\hhline{-||-||-|-}
\textbf{R18.3}  &            & \checkmark &            \\
\hhline{-||-||-|-}
\textbf{R18.6}  & \checkmark & \checkmark &            \\
\hhline{-||-||-|-}
\textbf{R19.1}  &            & \checkmark & \checkmark \\
\hhline{-||-||-|-}
\textbf{R21.13} &            &            & \checkmark \\
\hhline{-||-||-|-}
\textbf{R21.14} &            & \checkmark &            \\
\hhline{-||-||-|-}
\textbf{R21.17} &            & \checkmark & \checkmark \\
\hhline{-||-||-|-}
\textbf{R21.18} &            & \checkmark & \checkmark \\
\hhline{-||-||-|-}
\textbf{R21.19} &            & \checkmark &            \\
\hhline{-||-||-|-}
\textbf{R21.20} & \checkmark & \checkmark &            \\
\hhline{-||-||-|-}
\textbf{R22.1}  & \checkmark & \checkmark &            \\
\hhline{-||-||-|-}
\textbf{R22.2}  & \checkmark & \checkmark &            \\
\hhline{-||-||-|-}
\textbf{R22.3}  & \checkmark & \checkmark &            \\
\hhline{-||-||-|-}
\textbf{R22.4}  & \checkmark & \checkmark &            \\
\hhline{-||-||-|-}
\textbf{R22.5}  &            & \checkmark &            \\
\hhline{-||-||-|-}
\textbf{R22.6}  & \checkmark & \checkmark &            \\
\hhline{-||-||-|-}
\textbf{R22.7}  & \checkmark & \checkmark &            \\
\hhline{-||-||-|-}
\textbf{R22.8}  & \checkmark &            &            \\
\hhline{-||-||-|-}
\textbf{R22.9}  & \checkmark &            &            \\
\hhline{-||-||-|-}
\textbf{R22.10} & \checkmark &            &            \\
\hhline{-||-||-|-}
\end{tabular}
\begin{tablenotes}
\item[1] See Section~\ref{sec:MISRA-C-readability} for an alternative
         view on how to check compliance with respect to this rule.
\end{tablenotes}
\end{threeparttable}
\end{table}
}

The guidelines are listed in Table~\ref{tab:MISRA-C-semantic-guidelines}.
Note that the text provided for each guideline is just, as indicated,
a rough, very rough one-line summary, whereas
the proper description can span multiple pages.
The reader is referred to
\cite{MISRA-C-2012,MISRA-C-2012-Amendment1,MISRA-C-2012-TC1}
for the full details.
Note that the list of guidelines
in Table~\ref{tab:MISRA-C-semantic-guidelines}
begins with four directives: even though checking compliance with respect to
them requires information that may not be present in the code,
they involve undecidable program properties.

Table~\ref{tab:MISRA-C-semantic-guideline-properties} classifies the
guidelines of Table~\ref{tab:MISRA-C-semantic-guidelines} according to
attributes of an approximate representation of the program semantics;
an approximation built by a static analysis algorithm to check
compliance for the given guideline with adequate precision,
that is, no false negatives and relatively few false positives.
The attributes are the following:
\begin{description}
\item[control-flow:]
  detecting all potential violations with a low rate
  of false positives requires computing an approximation
  that allows observing control-flow within the program with
  relatively high precision;
\item[data-flow:]
  detecting all potential violations with a low rate
  of false positives requires computing an approximation
  that allows observing the possible values of objects with
  relatively high precision;
  this is further refined with two sub-attributes:
  \begin{description}
  \item[points-to:]
    observing the values of pointer objects is important;
  \item[arithmetic:]
    observing the values of other (i.e., non-pointer) objects
    (including pointer offsets) is important.
  \end{description}
\end{description}
Of course, it is well known that control-flow information depends
on data-flow information and the other way around, exactly as points-to
information depends on arithmetic values and the other way around:
here we only characterize the approximation
that is available \emph{at the end} of the static analysis,
without reference to how it has been obtained.

Table~\ref{tab:MISRA-C-semantic-guideline-properties} shows
that semantic-based static analysis potentially plays an important
role in the checking of compliance with respect to MISRA~C.
The actual situation, however, is not as clear cut:
this brings us to the next section.

\FloatBarrier

\section{Discussion}
\label{sec:discussion}

We have seen that~$40$ MISRA~C guidelines out of~$173$
depend on semantic properties of the program.
This implies that research in semantic static analysis
is very relevant to the MISRA~C ecosystem, provided that
a few important points are taken into due consideration.
These are discussed in Sections~\ref{sec:MISRA-C-vs-bug-finding},
\ref{sec:MISRA-C-readability}, and~\ref{sec:analysis-of-misra-c-code}.
A further opportunity for cooperation is outlined in
Section~\ref{sec:annotations}.

\subsection{MISRA~C: Error Prevention, Not Bug Finding}
\label{sec:MISRA-C-vs-bug-finding}

As said earlier, MISRA~C cannot be separated from the process
of documented software development it is part of.
In particular, the use of MISRA~C in its proper context is part
of an \emph{error prevention} strategy which has little in common
with \emph{bug finding}, i.e., the application of automatic
techniques for the detection of instances of some software errors.
This point is so rarely understood that it deserves proper explanation.

To start with, the violation of a guideline is not necessarily
a software error.  For instance, let us consider Rule 11.4, which
advises against converting integers to object pointers and vice-versa.
There is nothing intrinsically wrong about converting an integer constant
to a pointer when it is necessary to address memory mapped registers
or other hardware features.
However, such conversions are implementation-defined and have undefined
behaviors (due to possible truncation and the formation of invalid
and/or misaligned pointers), so that they are best avoided everywhere
apart from the very specific instances where they are both required
and safe.
This is why the deviation process is an essential part of MISRA~C:
the point of a guideline is not ``You should not do that'' but
``This is dangerous: you may only do that if (1) it is needed, (2) it is safe,
and (3) a peer can easily and quickly be convinced of both (1) and (2).''
One useful way to think about MISRA~C and the processes around it
is to consider them as an effective way of conducting
a \emph{guided peer review} to rule out most C~language traps
and pitfalls.\footnote{We are indebted to Clayton Weimer for
  this observation.}

The attitude with respect to incompleteness is entirely different
between the typical audience of bug finders and the typical
audience of MISRA~C.  Bug finders are usually tolerant about false negatives
and intolerant about false positives: for instance,
by following the development of
\emph{Clang Static Analyzer}\footnote{%
  \url{https://clang-analyzer.llvm.org/}, last accessed on July 5th, 2018.}
it can be seen that all is done to avoid false positives with little
or no regard to false negatives.
This is not the right mindset for checking compliance with respect
to MISRA~C: false positives are a nuisance and should be reduced and/or
confined as much a possible, but using algorithms with false negatives
implies that those in charge of ensuring compliance will have to use
other methods.
So, compliance to MISRA~C is not bug finding and, of course,
finding some, many or even all causes of run-time errors does not imply
compliance to MISRA~C.

\subsection{MISRA~C: Readability, Explainability, Code Reviews}
\label{sec:MISRA-C-readability}

Another aspect that places MISRA~C in a different camp from
bug finding has to do with the importance MISRA~C assigns to
reviews: code reviews, reviews of the code against design documents,
reviews of the latter against requirements.
Concerning design documents and requirements this is captured by
Directive~3.1.  More generally, the need for code readability
and explainability is clearly expressed in the rationale of many
MISRA~C guidelines.

This fact has some counterintuitive consequences
on the use of static analysis, which is of course crucial both for bug
finding and for the (partial) automation of MISRA~C compliance
checking.
Consider Rule~9.1, whereby the value of an automatic object
must not be read before it has been set, since otherwise we have undefined
behavior.  For bug finding, the smarter the static analysis algorithm
the better.
Use of the same smart algorithm for ensuring compliance with respect to Rule~9.1
risks obeying the letter of MISRA~C but not its spirit.\footnote{There
  are many ways to do that.}
Suppose on the specific program our smart algorithm
ensures Rule~9.1 is never violated: we have thus ruled out one
source of undefined behavior, which is good.
However, the programmer, other programmers, code reviewers, quality assurance
people, one month from now or six months from now may have to:
\begin{enumerate}
\item
  ensure that the automatic objects that are the subject of the rule
  are indeed initialized with the correct value;
\item
  confirm that the outcome of the tool is indeed correct.
\end{enumerate}
If this takes more than 30~seconds or a minute per object, this is not good:
the smart static analysis algorithm can track initializations and uses
even when they are scattered across, say, switch cases nested into
complex loops;  a human cannot.
So, ensuring compliance with respect to Rule~9.1 with deep semantic
analysis is counterproductive to the final goal of the process
of which MISRA~C is part.
For that purpose it is much better to use a decidable approximation
of Rule~9.1 such as a suitable generalization
of the \emph{Definite Assignment} algorithm employed
by Java compilers \cite[Chapter 16]{GoslingJSBB14}.

\subsection{Analysis of Code Meant To Comply with MISRA~C}
\label{sec:analysis-of-misra-c-code}

As was already recognized in \cite{CousotCFMMMR07}, despite the mentioned
misunderstanding about the nature of MISRA~C, the restriction to a language
subset where non-definite behavior and many problematic features
are banned or severely regulated ``can considerably
help the efficiency and precision of the static analysis.''
This can simplify and guide the design of static analyzers:
for example, features of C that are deprecated by MISRA~C need
not be handled precisely and efficiently when the intended application
domain follows MISRA~C.  It is not a coincidence that such features
(e.g., unions, unrestricted pointer casts, backward gotos) pose significant
problems to the designers of static analyses tools.

\subsection{Annotations}
\label{sec:annotations}

Another area where there is significant potential for collaboration
between the static analysis community and the MISRA~C ecosystem
concerns program annotations.
During the last 20 years there have been a number of proposals for
annotation languages allowing programmers to provide partial specifications
of program components.
These languages are usually tied to one specific tool, e.g.:
the annotation language of the
\emph{Frege Program Prover} \cite{Winkler97};
the annotation language of \emph{eCv} (Escher C Verifier) \cite{CrockerC07},
the annotation language of \emph{Frama-C},
\emph{ACSL} (ANSI/ISO C Specification Language) \cite{ACSL-1.13},
and its executable variant \emph{E-ACSL} \cite{EACSL-1.12};
the annotation language of \emph{VCC} \cite{DahlweidMSTS09};
the annotation language of \emph{VeriFast}
\cite{JacobsSPVPP11,PhilippaertsMPS0P14}.
A comparison of these tools, with particular regard to annotation
languages and the potential for application in industry,
is available in \cite{Rainer-Harbach11th}.

The MISRA~C Working Group is working, among other things,
on a tool-agnostic annotation language for {C}.
The main objectives of this endeavor are:
\begin{enumerate}
\item
  to improve the quality (precision) of static analysis by allowing the
  provision of information regarding the developer's intent, the
  required state in function preconditions and so on;
\item
  to make it easier to work with adopted code (legacy code, library code)
  that has not been written to comply with MISRA~C
  \cite{MISRA-Compliance-2016};
\item
  to do this in a way that will be accessible to the majority of
  C/\Cplusplus{} programmers in a form that is easy to read and understand.
\end{enumerate}

\section{Conclusion}
\label{sec:conclusion}

In this paper, having explained some of the advantages and
disadvantages of using the C language for embedded systems and how the
uncontrolled use of C conflicts with both safety and security
requirements, we described the background, motivation and history of
the MISRA project.  We have explained how the MISRA~C guidelines
define a standardized structured subset of the C
language; making it easier, for code that follows these guidelines
(possibly with well-documented deviations),
to verify that important and necessary safety and security properties hold.

We have looked at the different kinds of the MISRA~C guidelines,
distinguishing between those that can be automatically
verified from the code syntax,
those that need information beyond that contained in the source code,
and those for which the question as to
whether the code is compliant is algorithmically undecidable.
We have noted also that, for all guidelines, due to the size and complexity of
modern software, automatic tools perform an essential function in the
checking or partial checking of compliance.
We have highlighted the fundamental differences between so-called bug finding
and the application of MISRA~C in the context of the error prevention
strategy it is part of.

In this paper we have outlined both the role of static analysis
in the automatic checking of compliance with respect to MISRA~C, and
the role of the MISRA~C language subset in enabling a wider application
of formal methods to industrial software.
It is hoped that this will contribute to improved collaboration between
the two communities, so that
static analysis will be able to play a fuller part in the software development
of critical systems leading to improved safety and security.

\subsection*{Acknowledgments}

For the notes on the history of MISRA and MISRA~C we are indebted
to Andrew Banks (LDRA, current Chairman of the MISRA C Working Group)
and David Ward (HORIBA MIRA, current Chairman of the MISRA Project).
For the information on the ongoing work on annotations, we thank
Chris Tapp (LDRA, Keylevel Consultants, MISRA~C Working Group,
current Chairman of the MISRA~\Cplusplus{} Working Group).
We are grateful to the following people who helped in proofreading
the paper and provided useful comments and advice:
Fulvio Baccaglini (PRQA --- a Perforce Company, MISRA C Working Group),
Dave Banham (Rolls-Royce plc, MISRA C Working Group),
Daniel K\"astner (AbsInt, MISRA C Working Group),
Thomas Schunior Plum (Plum Hall, WG14),
Chris Tapp (ditto),
David Ward (ditto).
We are also grateful to the following BUGSENG collaborators:
Paolo Bolzoni, for some example ideas;
Anna Camerini for the composition of Figure~\ref{fig:misra-c-history}.

%\bibliographystyle{splncs04}
%\bibliography{misra,other}

\providecommand{\noopsort}[1]{}

\end{document}